\documentclass[manuscript]{aastex} % ** use this

\newcommand{\msun}{\rm M_{\odot}}

\slugcomment{Accepted by AJ}
\shorttitle{The Stellar Content of NGC 300}
\shortauthors{Butler et al.}

\begin{document}
\title{The Stellar Content and Star Formation History of the 
late-type spiral
  galaxy NGC 300 from
{\it Hubble Space Telescope}\thanks{Based on
  observations made with the 
  NASA/ESA Hubble Space Telescope, obtained from the data archive at the Space
  Telescope Science Institute. STScI is operated by the Association of
  Universities for Research in Astronomy, Inc. under NASA contract NAS
  5-26555.}   observations} 
\author{D. J. Butler, D. Mart\'{\i}nez-Delgado  \and W. Brandner}
\affil{Max-Planck-Institut f\"ur Astronomie, K\"onigstuhl 17, D-69117
  Heidelberg, Germany}
\email{butler@mpia.de, ddelgado@mpia.de, brandner@mpia.de}

\begin{abstract}
 We present the  first  WFPC2 V, I photometry for the Sculptor Group galaxy NGC
 300 in four fields ranging from the  centre to the outer edge. 
 We have  made the first measurement 
 of the star formation histories in two disk  fields: the oldest 
  stars were born at similar
  epochs and formation activity increased but at different mean rates.
 The main disk stellar population is predominantly old, 
 consisting of  RGB and AGB stars,  based on a synthetic colour
 magnitude diagram  analysis.  Z is found to have been more 
 metal poor than  0.006 (or 0.33\,Z$_{\odot}$) 
 with no evidence for significant change in the mean Z value
  over time in both disk  fields.
   In the central region, we find a dearth of bright stars 
  with respect to the two disk fields that cannot be explained
 by observational effects. Taken at face value, this finding would agree
 with the Davidge (1998) report of 
 suppressed star formation there during the past 10$^9$\,yr with respect to 
 his disk fields at larger radii; but the
  possibility  of significant 
 central extinction affecting our finding remains. 
 We have also determined the first distance modulus estimate based on the 
 tip of the red giant branch method: on the Cepheid distance scale of
 Ferrarese et al. (2000) we find   (m-M)$_\circ$ = 26.56$\pm$0.07 
 ($\pm$0.13)\,mag; and  a similar value from the Cepheid-independent
 empirical method by Lee et al. (1993), both  
 in good agreement  with the Cepheid distance determined 
 by Freedman et al. (2001).  A discrepancy between this and 
 the theoretical calibration of the red giant branch tip magnitude method remains.
 Finally, we report a newly detected   young   (up to about 10\,Myr) 
 stellar association of about average size ($\sim$ 140\,pc)
  in one of the disk fields. 
\end{abstract}

\section{Introduction}
 In recent years real observational evidence has emerged indicating
 that spiral galaxies    assembled
 their components (disk, halo, bulge) at different times and on  different
 timescales 
 (for example, Andredakis, Peletier \&  Balcells
 1995; Ibata et al. 2001, and references therein; 
 and Gratton et al. 2003, based on ages of 
 a small sample of three  Milky Way
 globular clusters). 
 Galaxy formation studies are usually performed using surveys; however,   
 it is  of obvious importance to take
 representative examples of Hubble types and to study them in detail.  
  In order to learn  about
  galaxy formation and evolution in detail, there are three basic diagnostic tools, namely
 morphology, kinematics and stellar populations.
 Of these, perhaps the most direct way of studying the evolution of a galaxy's 
  stellar component is with diffraction limited
 imaging using the Hubble Space Telescope (HST).
 
 NGC 300 is typical of a late-type spiral of type SA(s)d (Tully 1988) and 
 the  brightest of five main spiral galaxies that comprise the Sculptor group.
 With a Cepheid-based distance estimate of (m-M)$_\circ$ = 26.53 $\pm$ 0.07\,mag (Freedman et al. 2001) and its near face-on orientation, NGC 300 is well suited to studies of its stellar content. In the past decade,  many studies have been devoted to study the bright young stellar population in this galaxy. Consequently, there are now
 several signs pointing to recent massive star
   formation,  
   namely the presence  Wolf-Rayet stars (e.g. Schild et al. 2003) and  individual supergiants
   (Urbaneja et al. 2003; Bresolin et al. 2002a/b; Schild  \& Testor 1992; Humphreys 
et al. 1986; Massey et al. 1984),  young stellar  associations (Pietrzy\'nski
 et al. 2001), 
 Cepheids (Pietrzy\'nski et al. 2002),  H$_{\rm II}$
   regions (e.g. Soffner et al. 1996, 
 DeHarveng et al.  1988),  supernova remnants (Pannuti et
 al. 2000), and  X-ray emission (Read \& Pietsch 2001).
 The older stellar populations have also been targeted (see Davidge 1998 and
 references therein) 
 but high angular resolution imaging ($\la$0.05$^{\prime\prime}$) 
 is needed to probe this fainter stellar
   population. 
 
In this paper, we present the first study of the resolved old,
intermediate-age and young stellar population of the spiral galaxy 
 NGC 300 based on
colour-magnitude diagrams (CMDs) from HST observations. 
 The main motivation for optical
  imaging data is  sensitivity to a relatively wide range of temperatures in 
  non- or marginally extincted fields. This paper is composed as follows. 
 Data and data reduction are described in
  Sec.~\ref{obs_datared}. An overview of the stellar populations
 based on CMDs is given in Sec.~\ref{cmds}. 
 The distance measurement based on the tip  magnitude 
 of the red giant  branch 
 is described in  Sec.~\ref{TRGBdist}, and the 
 issue of internal extinction is discussed in Sec.~\ref{dust}.
 The  population of young stars is discussed in Sec.~\ref{youngstars}
 and the derivation of the 
 chemical enrichment and star formation history is presented in 
 Sec.~\ref{SFH}. Finally, we  discuss our findings in 
    Sec.~\ref{Discussion}, and summarize them in Sec.~\ref{Conclusions}.

\section{Data and data reduction}\label{obs_datared}

\begin{figure}[htp]
 \centering
 \includegraphics[width=15cm,height=15cm,angle=-90]{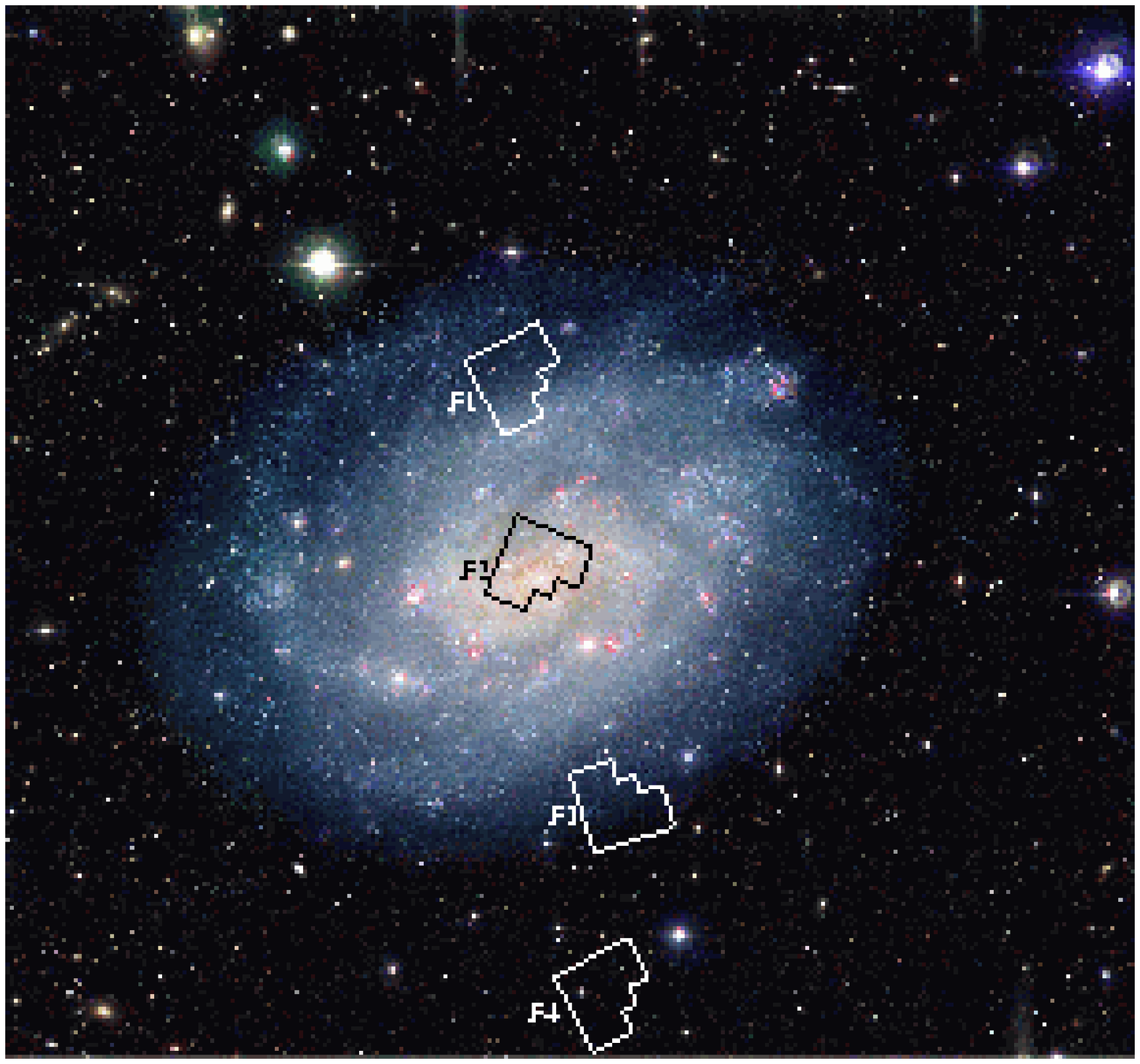} 
  \caption{European Southern Observatory/Wide Field Imager (colour composite)  image of NGC 300 with HST WFPC2 footprints 
  superimposed. The field size is 34.6$^{\prime}$   square.
   WFPC2 positions: r = 5.99$^{\prime}$ (F1); 0.44$^{\prime}$ (F2);
     7.12$^{\prime}$ (F3) ;  and
   12.84$^{\prime}$ (F4). North is up and east is left.
} \label{DSS2red_NGC300_Chart}
  \end{figure}

For this study, we retrieved the  
 calibrated science and data quality images of several NGC 300 fields from the Space Telescope  Science Institute HST data archive. We have examined  archive I$_{\rm 814}$- and V$_{\rm 547/555/606}$-band  frames from  four HST/WFPC2 
 pointings. These are listed in Table~\ref{obs_log}, and  their WFPC2 
footprints are superimposed on an image from the ESO Wide Fielder Imager 
 (hereafter referred to as ESO/WFI\footnote{See Pietrzynski et al. (2002) for
 observation details.}) 
  in Fig.~\ref{DSS2red_NGC300_Chart}. Data reduction details are given 
 by Bagget et al. (2002) and Holtzman et al. (1995). 

The photometry of the stars in NGC 300 was derived using the
 HSTPHOT (Dolphin 2000a)  point spread function (PSF)-fitting 
 photometry package which  is  designed for optimal reduction
 and analysis of WFPC2 data, and has been used recently in a number of stellar population studies (e.g. M{\'e}ndez et al. 2002, and references therein). The code utilizes a library of PSFs to account for PSF position dependence, and
  is optimized for the undersampled  PSFs
 present in WFPC2 data.  This package was also used to mask bad pixels and columns using the data-quality image and reject cosmic rays in the images.
  For transformation of  the WFPC2 photometry into the standard V-, I-band
 system, stellar magnitudes are calibrated by HSTphot using the charge 
 transfer efficiency and zero
 point magnitude corrections derived by Dolphin (2000b).
 The package is reported to be accurate to 0.02\,mag
 Dolphin (2000) with a 
 coordinate transformation  rms residuals  of the order of 
 0.02$^{\prime\prime}$ in both  axes.

Artificial star tests have been performed 
 using  HSTphot for the  V-band magnitude range 
 20 to 28 (mag) and V-I = -0.5 to 3 (mag) which are approximately the ranges 
 interest in our CMDs. 
 Each test consisted of choosing a star from  the V-band magnitude and colour
 range,  adding it to both the  V, and I-band images and
 then the image was reanalyzed by HSTphot.  In this way, the 
 effect of creating  additional crowding of stars has been minimized.  
 This was repeated   several thousand times for each WFPC2 frame.
 From all of the tests  a database of measurements for 
 different positions and input V-band magnitudes and colours is built.
 The ratio of the number of recovered
  stars to the number of added stars in a magnitude range indicates 
 the statistical probability 
 of recovering a star in that magnitude range. 
 Additionally, the database allows the photometry errors (recovered - input
  magnitudes) to be determined for each bandpass.

\begin{figure}[htp]
   \centering
\includegraphics[width=10cm,angle=-90]{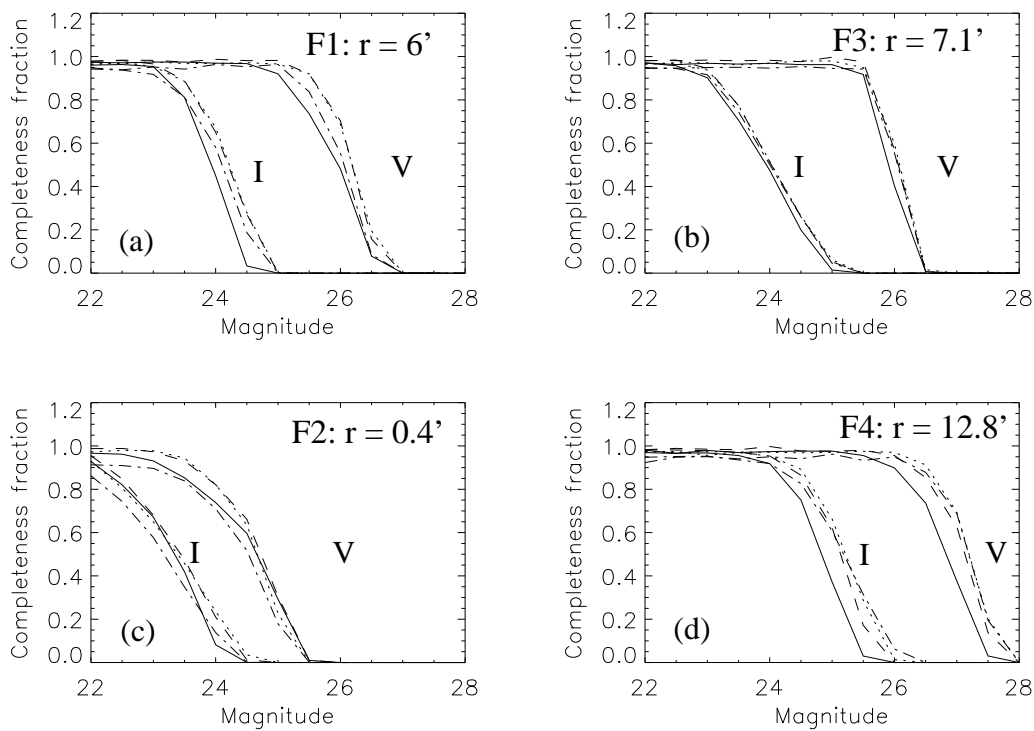}
   \caption{V, and I-band fractional completeness as a function of magnitude
    for WFPC-2   V-,  and I-band   PC1 ({\it solid}) , WF2 ({\it dotted}), WF3
 ({\it dashed}) and WF4 ({\it dot-dashed})
 frames  at each pointing as determined
  by artificial star tests. 
 Completeness 
 functions are the number of artificial stars recovered
  divided by the number of input stars; the bright ends of the functions in
 the above plots are
  typically less than  1.0 because of bad pixels.} \label{FigCompltnss}
    \end{figure}

  Fig.~\ref{FigCompltnss} plots the completeness factor  against magnitude
 for our four fields
 situated at different radial distances  from the  centre of NGC 300.   
 A completeness plateau occurs for the brightest stars because their
  signal-to-noise 
  is sufficiently high and crowding affects these stars equally. 
 A completeness of  100\% is not reached because of bad pixels. \footnote {The
 number 
 of pixels deemed to be bad 
  in the unvignetted portion  of each chip/frame is of the order of 1\% 
     (includes cosmic ray events;  and includes columns and pixels flagged as 
 bad in the STScI data quality file).}  
  For each band pass, the completeness is different 
 at  each pointing.
 The F4 field  has the longest integration time, providing
 the faintest magnitude limit and the crowding factor is shifted accordingly to fainter magnitudes relative to 
 the other integrations.
 There is a strong difference between the functions of 
 the  innermost and outermost fields. 
 Indeed, detectability in 
 the inner field is less efficient than in the outer fields
 largely due to severe crowding, 
 as well as shorter integration times. 
 The V-band functions have fainter limiting magnitudes than in I-band
  largely because of lower HST/WFPC-2 throughput in the I$_{\rm 814}$-band.
 
For the final photometry list used in this study, we select those objects flagged as valid stars in both band-passes and have  S/N $>$ 5;
$\chi^2$ $<$ 5; -0.5 $<$ sharpness $<$ 0.5; and $\sigma_{\rm V, I}$ $<$
0.25\,mag.   After selection, there are 7614, 1421, 3467 and 235 stars in the 
  F1, F2, F3 and F4 fields respectively.
 In the present paper we apply the selection criteria 
 to stars at I $<$ 23\,mag whenever star counts in colour magnitude
 diagrams are analyzed.
 At this magnitude limit  and brighter, the  artificial
 star tests find the  data to be  at least
  90\% complete,  and stars intrinsically brighter than I = 23\,mag
 will only be missed if they are physically obscured, e.g. by dust.

\section {The stellar content of NGC 300 from the colour-magnitude diagram}\label{cmds}

\begin{figure}[htp]
\includegraphics[width=15cm,height=14cm,angle=-90]{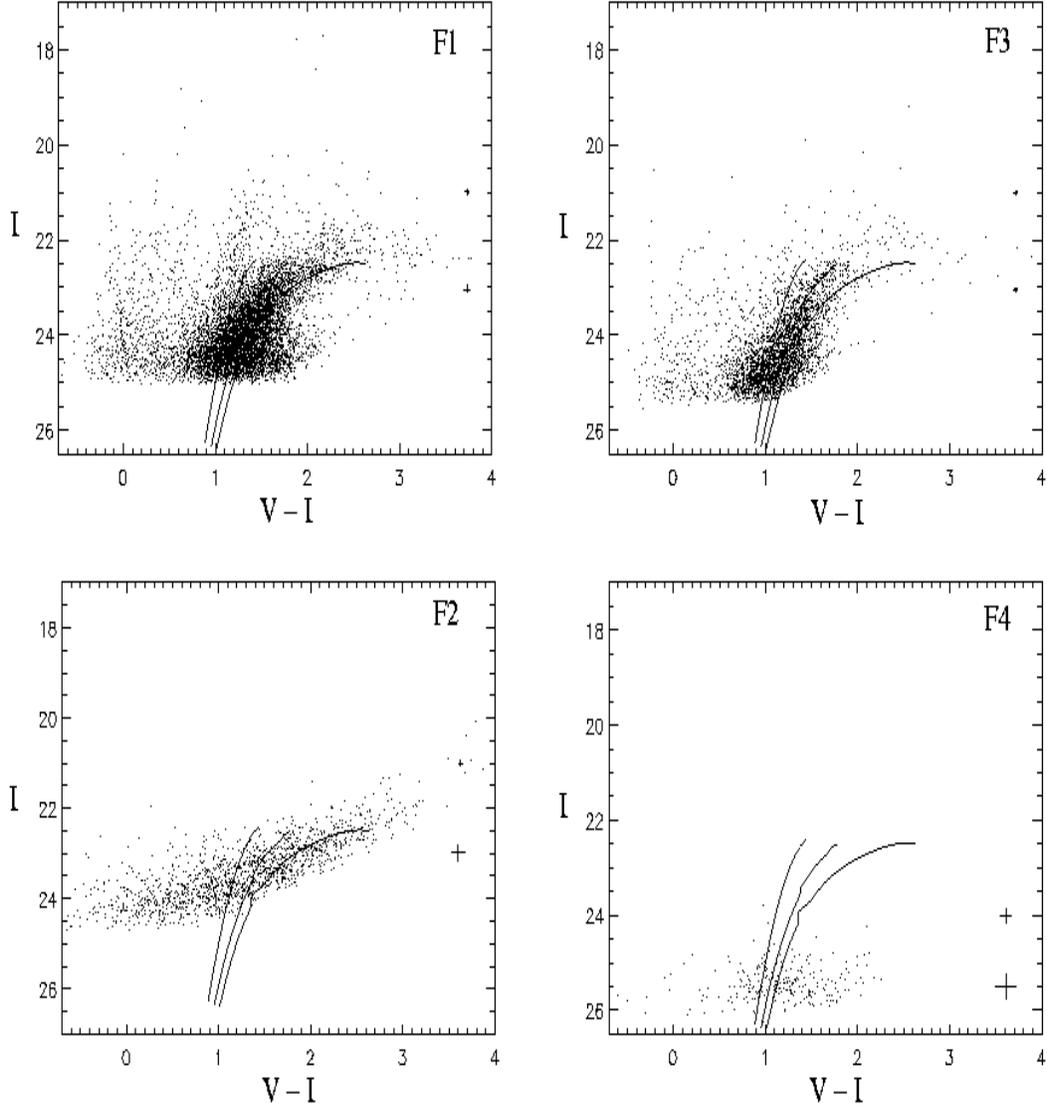} 
 \caption{I, V-I CMD for four WFPC2 pointings in
  NGC 300.  The plotted 
 data consists of  all stars with valid photometry (see 
 Sec.~\ref{obs_datared}). 
 There are three globular cluster red giant branch
 fiducial  ridgelines ({\it solid}), reddened using the A$_{\rm V}$
 and  A$_{\rm I}$ values given in Table~\ref{dist_details}. 
 From metal-rich (right) to metal poor (left), 
 they are for 47 Tuc, [Fe/H] = -0.71; NGC 1851, 
 [Fe/H] = -1.29;  and NGC 6397, [Fe/H] = -1.91. Included are  photometry
  bars associated  with V - I = 1\,mag.
} 
  \label{FigVIcmds}
 \end{figure}

Fig~\ref{FigVIcmds} shows the CMDs for the four WFPC2 fields described in
Table~\ref{obs_log}. These fields occur at radial distances ranging from
0.44$^{\prime}$ to 12.8$^{\prime}$ or de-projected radial distances of 0.4 to
10.9\,kpc at the distance of the NGC 300 for a distance of 2.02\,Mpc$\pm$0.07
(Freedman et al. 2001).

The  F1 and F3 fields overlap with parts of  different spiral arms in NGC 300
 (see Fig~\ref{DSS2red_NGC300_Chart}) and the CMD of each region  provide
 a sketch of some of the stellar content there. The
most prominent feature is the red giant branch (RGB) structure, which is very
similar to those observed in nearby dwarf irregular galaxies (NGC 6822:
Gallart et al. 1996; Sextans A: Dohm-Palmer et al. 2002). This structure is
typical of a galaxy that is clearly composed of an ancient
 stellar population: it is the locus of old
and intermediate-age RGB stars, low asymptotic giant branch (AGB) stars and
 blue-loop stars,  some hundred million years old. 
A significant number of red bright stars are observed
 above the tips of the fiducial RGB ridgelines and they  could be
 intermediate-age AGB stars covering a wide range of ages and
 metallicities. In addition, there is also an important population of young
 stars (age $<$ 1\,Gyr): these include main sequence stars;  core helium
 burning stars in the blue supergiant and/or in the blue loop region  at V-I
 $\la$ 0.9\,(mag); and red super-giants  at 0.9$\la$ V-I (mag) $\la$ 1.3. 

\begin{figure}[htp]
\includegraphics[width=15cm,height=14cm,angle=-90]{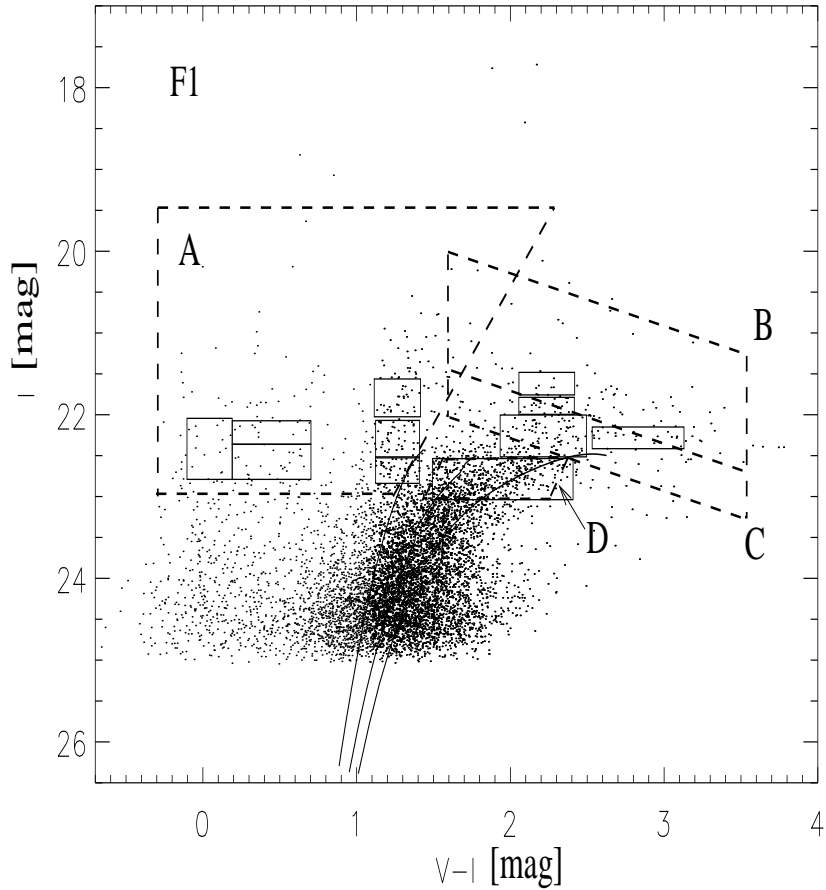}
 \caption{I, V-I CMD for field F1 including the 
 four labeled regions used for the star count analysis
 in Sec.~\ref{cmds} ({\it dashed}).
 Also included are the 
 eleven CMD boxes used to constrain the synthetic CMD ({\it solid}); see text in Sec.~\ref{SFH} for further details. The same
     boxes have been used for both the F1 and F3 fields. For the purpose of 
 convenience,  the stellar isochrones   described in  the caption of 
 Fig.~\ref{FigVIcmd_isoc} are also included.
} 
  \label{Star_XY_cmd_boxes}
 \end{figure}

The CMDs in these different regions also display  some marked differences in
 morphology that cannot be not fully explained by differences in observational
 effects. Most notably, the CMD of the innermost region appears stretched to
 red colours, mainly due to severe stellar crowding as judged by the
 completeness tests of the previous section. Incidentally, this  crowding
 issue highlights the need for near-diffraction limited imaging at extremely
 large ($>$ 20\,m class) optical/IR telescopes.   We see that there is an  absence of stars at  I$<$ 22\,mag in field F2
  that is not explained by incompleteness; the significance of this is
  discussed later  in Sec. ~\ref{dust} and Sec. ~\ref{youngstars}. 
 Lastly, we see a  dense string of stars at 
 V-I $\sim$ 0 and I$<$ 23\,mag in the F1 field that could well be the stellar 
 main sequence; the finding is especially noteworthy because 
  incompleteness  would tend to suppress such features 
 (see Fig.~\ref{Star_XY_cmd_boxes}).

For a comparative analysis of the star counts in the 
  fields at intermediate radii (F1 and F3),  
 we determined the ratio of star counts in three 
 age-sensitive CMD boxes  to the  star counts in an RGB/old AGB
 box.  The boxes are indicated in Fig.~\ref{Star_XY_cmd_boxes} and 
 the associated star counts and ratios are given  in 
 Table~\ref{table_F1F3counts}. 
 The data indicates that the F1 field has on average 3 - 4 
 times more stars, or similarly, that it is brighter by
 1 - 1.5 magnitudes than the F3 field. We also find that 
 there is a young population gradient
 in the F1 field, probably because it covers
 a possible spiral arm star forming region (Fig.~\ref{Star_XY}). 
 Lastly, unlike the old populations ($>$ 1\,Gyr),  
 there is a marked difference in the relative number of  
 young stars in the two fields. 
    In the outermost field (F4) there is a stark absence of  bright 
 stars (I $\la$ 24\,mag)  with respect to the F1 field.   
 As the surface  brightness difference  with respect to the F1
 field is  significant ($\sim$ 2.7\,magnitudes)    
 as measured by Carignan (1985), our observation of a dim disk field
 would be  expected.

The expected number of foreground stars 
 contaminating   our CMD was estimated based on the Kim et al. (2002)
 study, galactic models, and the Hubble Deep Field
 to be about 1$\pm$1 and about 5 
 per pointing, at I$<$17.5\,mag and I$<$23\,mag respectively. The number of 
 contaminating galaxies was estimated for several 
 fields of the Medium Deep Survey (Griffiths et al. 1994) -- 
 and our selection criteria (stars brighter than I=23\,mag) is expected 
 to be robust against background galaxy contamination at these bright
 magnitudes, estimated to be 1$\pm$1  per
 WFPC2 pointing and thus an unlikely source of contamination which
 was not   assessed by the  artificial  star test.
  The expected number of (unobscured) globular clusters in the WFPC2 fields 
   based on  Kim et al. (2002; hereafter
 referred to as K02) is at most 1$\pm$ 1.  Such objects would have
  V-I colours mostly in the range 0.9-2.0, and integrated I-band magnitudes of
 about 16-17\,mag at the distance of NGC 300 based on the Milky Way globular
 cluster system (Harris 1996).\footnote{We have checked whether some of the candidate  globular clusters from Kim et al. (2002) appear in the WFPC2 frames. One of them (their ID 8) occurs in the F1 field (chip 1)  while the other (ID 6) 
 appears in the centre field (chip 1). However,  both objects are saturated 
 in
 the WFPC2 frames, and there are neighbouring stars,
  further study of them has not been pursued.}

\section{TRGB distance to NGC 300}\label{TRGBdist}
The most  recent estimate of the distance to NGC 300 is by Freedman et
 al. (2001) using Cepheids.
  In the context of distance verification, it is important to verify this
 estimate using a different  distance indicator.
 We have used the  dereddened 
 magnitude of the TRGB, I$_{\rm \circ, TRGB}$,   to derive
  the distance modulus of NGC 300
 based on the  widely used 
 approach described by Lee, Freedman, \& Madore (1993, hereafter LFM93), and 
 typically has  errors less than 10\%. It
 has been assessed theoretically by Salaris \& Cassisi (1998; hereafter SC98)
  who found 
 that differences between TRGB and 
 Cepheid distances are no correlated with metal content.
  Incidentally, it is the first such
 distance measurement for NGC 300. 
 Distances to several other  Sculptor Group galaxies have been 
 measured previously using the same method (Karachentsev et al. 2003). 

   For the I$_{\rm TRGB}$ determination we consider the F1 and F3 
 fields, and ignore the remaining two fields due to a low number of stars and 
 blending in the innermost field.
  To reduce  the contribution of young/intermediate age stars, 
  we selected stars with (V-I) $>$ 1.5 (mag). 
  To measure the I-band magnitude of the TRGB, we applied the
 edge detection method described by Sakai, Freedman \& Madore (1996). 
 Firstly, for a sequence of magnitude values,  we created  
 a smoothed luminosity function, $\Phi$(m), by replacing each 
 magnitude value by a Gaussian distribution whose width (standard deviation)
  is the magnitude
  uncertainty (e.g. Fig.~\ref{Irgbt_N6E1} (panel 3)). The edge-detection response is 

\begin{equation}
\rm ED = \Phi(\rm m - \bar{\sigma_{\rm m}}) - \Phi(\rm m + \bar{\sigma_{\rm m}} ) 
\end{equation}

 where $\bar{\sigma_{\rm m}}$ is the average $\sigma_{\rm m}$ at
 m$\pm$0.05\,mag. ED was calculated
 at I-band magnitude  steps of 0.01\,mag.
 Next, we took 85\% of these stars in a random way, 
 and as a first
 estimate of the I$_{\rm TRGB}$ magnitude we recorded the 
 magnitude at which the filter response is a maximum at I $<$ 22.7\,mag.
 To determine the uncertainty in 
 the tip magnitude, we 
   performed a so-called bootstrap resampling of the data 
 by  repeating the procedure 500 times. 
 We then fitted a Gaussian
 to a histogram of the tip magnitude estimates  (binned at 0.01\,mag). 
 The peak magnitude from the Gaussian fit is taken as  I$_{\rm TRGB}$ and the 
 error in the Gaussian is taken as 
 the uncertainty (Fig.~\ref{Irgbt_N6E1}, top).
 These values  are given in Table~\ref{dist_details}. 
 For the F3 field, the tip magnitude estimate is 
 not reliable due to the low number of star counts, as
 is seen by the poor convergence of tip estimates
   (Fig.~\ref{centre_offIrgbt}, top).

 To determine how much these values depend on 
  the fraction of stars selected, 
  we varied the fraction selected from 0.6 to 0.9, and got the same
  I$_{\rm TRGB}$ value and only a slightly larger  
 uncertainty. 
 To test for a possible  bias on the TRGB magnitude caused by
  incompleteness we multiplied the smoothed luminosity function by
 a monotonically decreasing function  [1.0:0.80] in the magnitude
 range 22-24\,mag and found no sensitivity to this. 
  Another source of potential bias  comes from  
 smoothing the luminosity function. Cioni et al. (2000) reported the Sobel 
  filter to be a biased estimator 
 of the TRGB and that the bias depends on the amount of 
 smoothing in the luminosity function at the location of the TRGB. 
 To test our sensitivity to smoothing we varied the smoothing window
 in the range 0.04-0.06\,mag and find negligible change in the mean tip
  magnitude and a slight  change in errors.

The  absolute I-band extinction, A$_{\rm I}$,  toward NGC 300 can be derived  assuming
 the reddening law  R$_{\rm V}$ =[A$_{\rm V}$/E(B-V)] 
 = 3.1  and A$_{\rm I}$/A$_{\rm V}$ = 0.48 (Cardelli, Clayton \& Mathias 1989).
 From the COBE/DIRBE and IRAS
 dust maps analysed by  Schlegel, Finkbeiner, \& Davis (1998; hereafter
 referred to as SFD98) we have
 E(B-V) =  0.013\,mag at its galactic coordinates  (l = 299.21\,deg, b=
 -79.42\,deg from Simbad), which is similar to the 0.008\,mag from the lower resolution 
  dust maps of  Burstein \& Heiles (1982; hereafter referred to as BH82).
 Adopting the SFD98 value, we  get A$_{\rm V}$ = 0.039\,mag, and
  A$_{\rm I}$ = 0.019\,mag.
 The values are included in the dereddened I-band magnitude of  
  the TRGB  in  Table~\ref{dist_details}.  
 The possible presence of significant amounts of dust in the observed fields
 is a  concern that is explored in more detail in Sec.~\ref{dust}.

 Next, we calculate  M$_{\rm I, TRGB}$, using three methods for comparative
 reasons. In the first method 
 we adopt the value M$_{\rm I, TRGB}$ =-4.06$\pm$ 0.07
 (random) $\pm$ 0.13 (systematic)\,mag
 from Ferrarese et al. (2000; hereafter referred to as F00). They treated the
 TRGB as a secondary distance indicator and calibrated a zero point from
 galaxies with  Cepheid distances. 
  The  calculated distance modulus, (m-M)$_{\rm \circ}$ is, therefore, 
  26.56   $\pm$0.07 (random) $\pm$0.13 (systematic) \,mag.
 Importantly,
  it is in good agreement with the recent 
 Cepheid-based estimate by Freedman et al. (2001).
 The caveat is that the present
  distance estimate is not  strictly an independent
 empirical estimate. 

 The second method is the empirical, Cepheid-independent one   
 suggested by Lee et al. (1993). Summarized by SC98 (their p\,168), it 
 is based on the relation between the (dereddened)
  I-band distance modulus and  
  I$_{\rm \circ, TRGB}$, i.e. \rm (m - M)$_{\rm \circ, I}$ = I$_{\rm \circ, TRGB}$ + BC$_{\rm I}$ - M$_{\rm bol, TRGB}$.
 The I-band bolometric correction, BC$_{\rm I}$, is determined using 
\begin{equation}\label{eqn2}
\rm BC_{\rm I} = [-0.881 - 0.243(\rm V-I)_{\rm \circ, TRGB}] \pm 0.057\,mag
\end{equation}
 from Da Costa \& Armandroff (1990) where $(\rm V-I)_{\rm \circ, TRGB}$ is 
 the (dereddened) colour of the RGB locus at the tip magnitude.
  M$_{\rm bol, TRGB}$ is determined using 

\begin{equation}\label{eqn3}
\rm M_{\rm bol, TRGB} = -0.19\,[Fe/H] - 3.81, 
\end{equation}
 and [Fe/H]  of the parent stellar population is related to the (dereddened) colour of the RGB locus at M$_{\rm I
 }$ = -3.5\,mag ((V-I)$_{\rm \circ, -3.5}$)  via  
\begin{equation}\label{eqn4}
\rm [Fe/H] = -12.64 + 12.6\,(V-I)_{\circ, -3.5} - 3.3 \,(V-I)_{\circ, -3.5}^2
\end{equation}
both from Lee et al. (1993). 
 With  $(\rm V-I)_{\rm \circ, TRGB}$ = 1.82$\pm$0.1\,mag\footnote{Colour
 values were derived in a bootstrap way:  
 85\% of the stars were randomly selected in an I-band interval of 0.1\,mag
  (i.e. [I$_{\rm TRGB}$ : I$_{\rm TRGB}$ + 0.1\,mag] in the case of the RGB
 tip);  
 and we then repeated this 
 500 times in order to obtain the average estimate with the 
 standard deviation of the fit to the peak of the resultant
  data (binned at 0.05\,mag)
  taken as the uncertainty.} 
 and (V-I)$_{\rm \circ, -3.5}$ = 1.58\,dex (field F1)
 we find [Fe/H] = -1.05\,dex and 
  (m - M)$_{\circ}$ = 26.62\,mag $\pm$  0.06 (random; from I$_{\rm TRGB}$). 
This is in good agreement
 with the Cepheid-based calibration.

In the third method, we calculated M$_{\rm I, TRGB}$ using the theoretical
 calibration of
 SC98: the recipe  makes use of their re-calibrated
 relation between the global metal-to-hyrodgen ratio 
 [M/H] and  (V-I)$_{\rm \circ, -3.5}$:
\begin{equation}
\rm [M/H] = -39.270 + 64.687\,[(V-I)_{\circ, -3.5}] - 36.351\,[(V-I)_{\circ,
  -3.5}]^2 + 6.838\,[(V-I)_{\circ,  -3.5}]^3.
\end{equation}
 The SC98 recipe
  is based on updated stellar models, the empirical 
 calibration of synthetic colours,
 and the adopted bolometric correction. They obtained the following
  relation  for the absolute I-band   magnitude of the TRGB:
\begin{equation}
\rm  M_{\rm I, TRGB} = - 3.953 + 0.437\,[M/H] + 0.147\,[M/H]^2
\end{equation}
 which leads to [M/H]=-0.85\,dex, and 
 (m-M)$_\circ$ =    26.73 $\pm$ 0.06\,mag (random). 
 The  difference between this value  and the value from the 
 Cepheid calibration (i.e. 26.56\,mag; Freedman et al. 2001) is 0.17\,mag.

 The good agreement of the first method (empirical TRGB calibration of the
 distance tied to the Cepheid scale; F00), with 
 Cepheid distance itself is not surprising. The TRGB distance 
 from the Lee et al. (1993) calibration is an independent 
 measure of the distance to NGC 300 and is in good agreement
 with the F00 measurement. There is, however, a discrepancy 
 with the theoretical calibration of SC98, which may be a zero-point
  problem.\footnote{
 However, in support of the F00 calibration, we 
 note that that a recent  RR Lyrae star calibration of the LMC distance
  modulus by Clementini et al., (2003) gave
 18.515 $\pm$ 0.085\,mag which is in good aggreement with  the value
 (18.50\,mag)  adopted by F00.} 
 In conclusion, we  adopt the F00 value of M$_{\rm I, TRGB}$
 because it is  precise and most especially because   
 it is a statistically averaged solution whose zero-point
 derivation is based on a homogeneous and consistent calibration.
 The corresponding  distance modulus estimate  is given in 
 Table~\ref{dist_details}.

\section{Internal Extinction in NGC 300}\label{dust}

For our study of the bright stellar populations of NGC 300, it is very
 important to consider the issue of  internal reddening
 of the galaxy because it can broaden CMD features:   internal
 extinction may vary locally depending on the spatial distribution of dust and
 stars (e.g. Jansen et al. 1994). It is   
 probably low on average based on 
 the dust maps of SFD98, which are in fact in good agreement with the values 
 from the lower resolution  dust maps of BH82 as mentioned in the previous 
 section.  Going one step further,  we looked for a difference in  extinction
  between the F1 and F3 fields.  From the previous section,  we know that the
 TRGB magnitudes in these fields appear 
 not to differ by more than about 0.1\,mag at most,   
 which argues against significant 
 differential extinction in these fields. This is especially meaningful
 because the I-band TRGB  magnitude 
 is fairly insensitive to  variations in heavy element content (Salaris, 
 Cassisi, \& Weiss 2002), something that
  could partially mimic differential extinction.
 
Another issue is extinction in the nuclear region of NGC 300. Indeed, although 
  NGC 300 is  near face-on, as opposed to the  edge-on 
 case in which central extinction  tends to be higher,
 there is the possibility of 
 significant extinction in the central region (Peletier et al. 1995). 
 In contrast to the outer fields,
  there are clear shells of dust in the central region based on visual
 inspection of the ESO/WFI (real-) three-colour image \footnote{A high resolution colour image is included in the online electronic version}
  (Fig.~\ref{DSS2red_NGC300_Chart}). These shells 
 are distributed in a  fragmented, ring-like structure 
 of de-projected diameter $\sim$ 1.2\,kpc ($\sim$  1$^{\prime}$), 
  with an annular thickness of 0.2-0.4\,kpc ($\sim$  10-20$^{\prime\prime}$),
  based on visual inspection.
 The dust ring appears to be symmetric about the 
  bright compact central object  studied recently by B\"oker et al. (2002). 
 Significantly, there is a  stark absence of blue 
 colours  in the image  which suggests that there may be significant 
 extinction, possibly hiding evidence of recent star formation, 
  and /or that there has been relatively little 
  recent star formation  in the nuclear and circum-nuclear region.
  For the central region, Davidge (1998) found evidence to support the idea of 
 suppression of recent star formation, a plausible result because the 
 study  was based on  essentially extinction-free near-IR imaging.

 Lastly, we remark that 
 almost no reddening has been reported for two early-type super-giants, one
 near the galaxy centre (at r  $\sim$ 0.55\arcmin)
 and one in the outskirts (at r $\sim$ 9.16\arcmin) (Urbaneja et
 al. 2002).  However, it is very difficult to make
 specific conclusions regarding
 field-to-field differences in extinction from such a sparse sample and
  a larger sample is needed in order to place wide-field spectroscopy-based
 reddening information on a firm statistical footing.

\section{The Young Stellar Population}\label{youngstars} % \label{Z_discuss}

\begin{figure}[htp]
 \includegraphics[width=8cm,height=9cm,angle=-90]{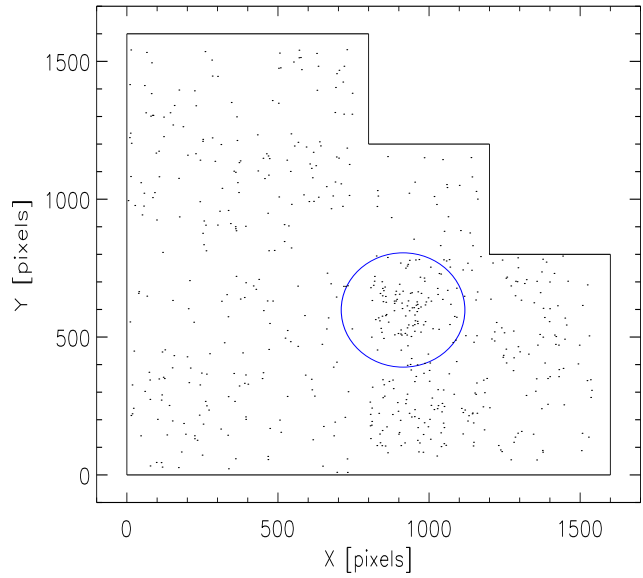} \\
\includegraphics[width=9cm,height=9cm,angle=0]{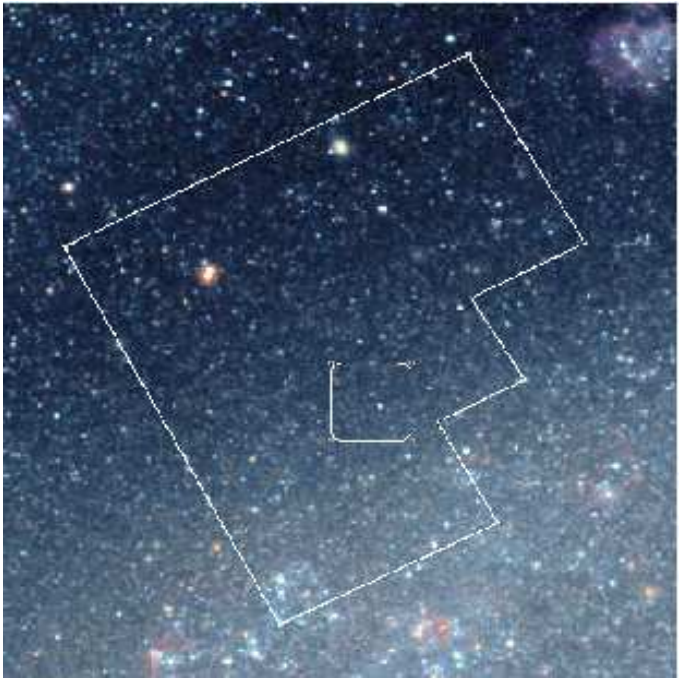} 
  \caption{Star positions in the F1  field for stars younger than
     1\,Gyr (top) and section of the ESO/WFI image with the WFPC2 FoV
 overlaid.
 The discontinuity at columns 750 to 800\,pixels (top)
 results from our setting their values to  
  zero because of  vignetting by the WFPC2 optics. In the ESO/WFI field, 
  north is up and east is left. The small box  indicates the approximate
 position of the newly detected stellar association. 
  See text in Sec.~\ref{youngstars} for further  information. } \label{Star_XY}
 \end{figure}

\begin{figure}[htp]
\includegraphics[width=13cm,height=15cm]{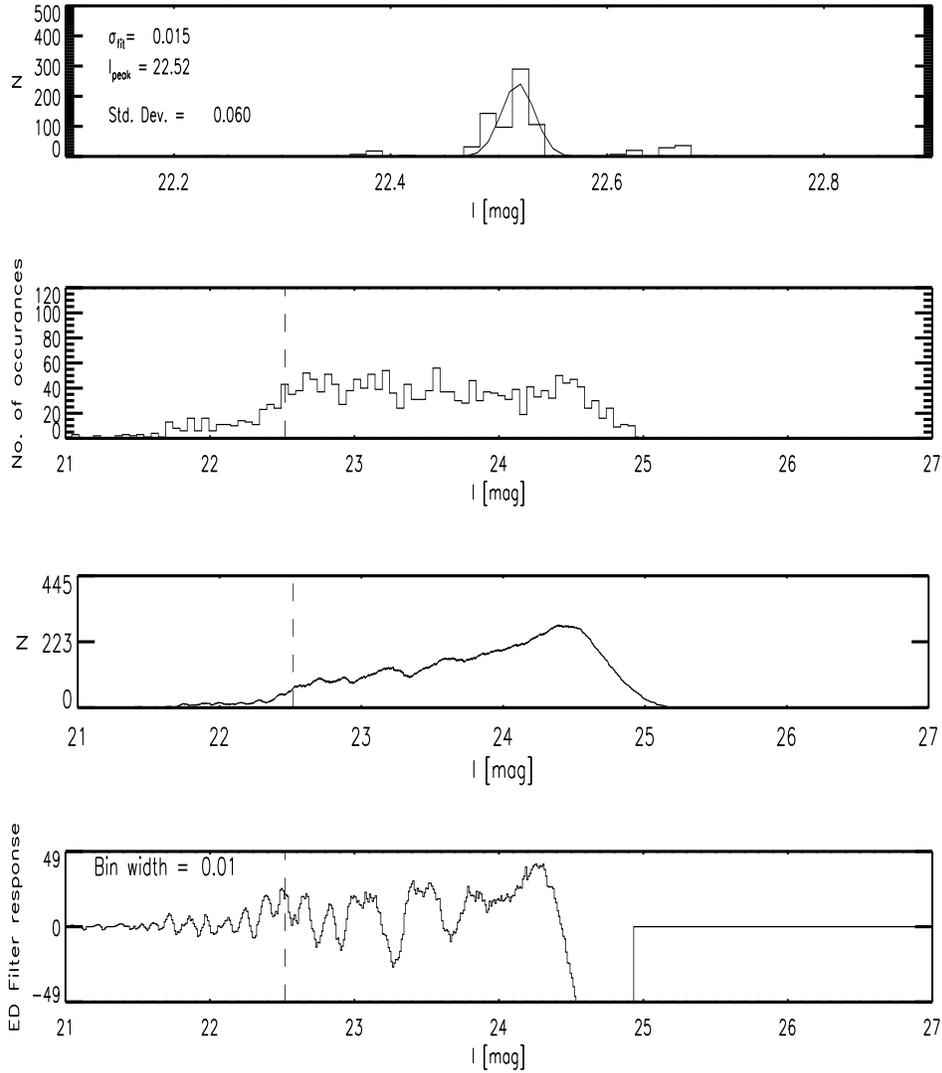}
   \caption{Deriving RGB tip magnitude in the F1 field.
   Top to bottom: Histogram of derived TRGB tip magnitudes, histogram
 version of the luminosity function, smoothed luminosity function, and 
 an example  edge  detection filter response. }  \label{Irgbt_N6E1}
    \end{figure}

\begin{figure}[htp]
\includegraphics[width=13cm,height=15cm]{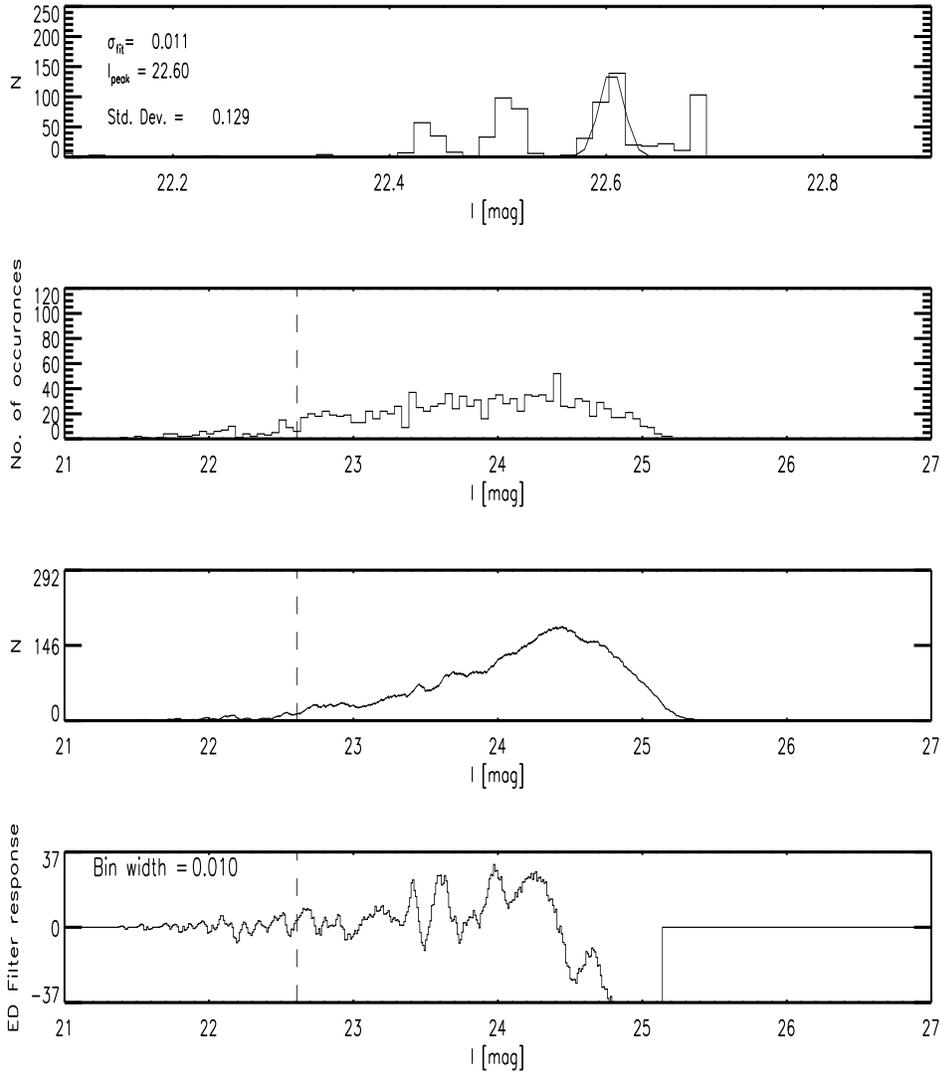}
   \caption{Deriving the RGB tip magnitude  for the F3 field. Panels are 
    as in Fig~\ref{Irgbt_N6E1}. 
Note that there is no clear unimodal-like distribution of bootstrap
     estimates in the top panel.}\label{centre_offIrgbt}
    \end{figure}

 We checked whether  there are any spatially distinct stellar 
 groups or associations of stars younger than about 1\,Gyr
 in the F1 and F3 fields at I$<$ 24\,(mag) (i.e. an extension of box A in  
 Fig.~\ref{Star_XY_cmd_boxes}). 
 For this, a histogram of the number density of stars in 100 pixel wide
 columns   was then computed.
 In each  histogram, a background and 
  4$\sigma$ detection threshold level was  determined 
 from the first 750 columns of the WFPC2 frame.
 In this way, any  detection is  
 neither affected by the WFPC2 L-shape, shown in  Fig.~\ref{Star_XY},
   nor statistical fluctuations. 
 While there is no real evidence for a stellar group in the F3 field, 
 there is a significant group in the the other field (F1). 
  For validation of the detection,
 we varied bin size from 
  50 to 150\,pixels; and the significance remains.
  Based on a  Gaussian fit to the histogram for field F1, we find
 the size (FWHM) 
 to be about 140\,pc, 
  at a distance of 2.02$\pm$0.07\,Mpc
  (e.g. Freedman et al. 2001). This is consistent with the average 
  OB association size in NGC 300 (Pietrzy\'nski et al. 2001).  
  Based on the  work of Kim et al. (2002) who age-dated 
  15\% of the 117 OB associations cataloged by Pietrzy\'nski et
 al. (2001), the  age of the newly detected  association 
 would be of the order of  10\,Myr or less, and is approximately
 of average size. Accordingly, bright main sequence 
 and/or blue supergiant stars may be present.
  We note that none of the X-ray sources  detected by the ROSAT HRI
  (Read et al. 2001) occur in the  F1 field, or the other fields.
   The coordinates of the association are given in
  Table~\ref{table_newAssociation}.  

 Another issue is star forming environments.
 The F1 field covers parts of a major and minor spiral arm 
 (see Fig.~\ref{DSS2red_NGC300_Chart} and Fig.~\ref{Star_XY} (bottom)).
 The blueness of these arms  with respect to the interarm region 
 argues for a rich supply of young stars at the southern end of the 
 F1 field which is also indicated by  Fig.~\ref{Star_XY} (top).
 Consequently, it is not suprising that there are parts of 
OB associations cataloged by Pietrzynski et al. (2001), 
 namely AS 054 and AS 055, at the southern end of the WFPC2 FoV
(Fig.~\ref{Star_XY}). 
 We note that of the 
 Wolf-Rayet stars found by  the Schild et al. (2003) survey of several fields 
  and the  associations of OB stars cataloged by Pietrzy\'nski et al. (2001), 
  a few such stars  occur in our fields, 
  but did not meet our selection criteria (see Sec.~\ref{obs_datared}:
 i.e. rejected as saturated; stellar blends; and 
  or large fitting/photometry errors)  and have been ignored.

  Pursuing the SF issue further, 
 we refer to cataloged blue stars,
  OB associations, and the correlation made with H$_{\rm II}$ maps by
Pietrzy\'nski et al. (2001). 
  From their  Fig. 4  we see that there is a significant
 drop-off in the number of OB associations at r $\sim$ 9-12${^\prime}$
  (projected).  Consequently, if these were used as tracers of  star
 formation,  one would conclude that such activity  is significantly lower
 at large radii with respect to that of the disk.
 We also note that there are some  blue stars
 at r = 12-13.5$^{\prime}$ that might be tracing the galaxy out to even larger
 radii, or possibly associated with parts
  of tidal streams;  but we cannot exclude the  possibility  
 that such stars are simply just the bright  field star population that one
 expects  (see  Sec.~\ref{SFH}).

 The next issue is the amount of recent star formation.
   Based on the presence of blue helium burning stars
  in Fig.~\ref{FigVIcmd_isoc},
 a marked recent burst of star formation in the observed fields is
 likely. However, the small WFPC2  field of view certainly
  means that the brightest population of stars may  not have been be sampled
  adequately. Nevertheless, we can say 
 that star formation may well be continuing at the observed
   pointing, statistically speaking, based on  the
  presence of  a   blue plume 
 stretching up to V$\sim$18\,mag  (I$\sim$18\,mag also)
  in the (V,B-V) CMD of  Pietrzynski et al. (2001)  which 
 covers the  whole galaxy. Such stars may be in OB associations, or 
 less concentrated star forming regions. 
 Guided by the isochrones in the same figure, 
 we also see clues there may have been very recent star formation
 because of some bright,  
 possibly zero-age main sequence stars at V-I$\sim$-0.4\,mag that are present. 
 
 Lastly, there is evidence for a difference in the stellar content of the field F1 and
 F3. One can  see that there is a relative  dearth  of  intermediate 
 mass stars in the range 5-12\,M$_\odot$ in the F3 field
 with respect  to the F1 field:  
 such stars in the F1 field are predominantly associated with
  the spiral arms (see Fig.~\ref{Star_XY}).
 We recall from Sec.~\ref{cmds} that the ratio of old 
 to young ($\la$ 1\,Gyr)  stars for the two fields are not in good agreement
  (see Table~\ref{table_F1F3counts}),  and so
  the observation cannot be attributed completely to the difference in 
 surface brightness  (or the mean star formation activity) between  
 two disk fields (see Sec.~\ref{cmds}). It can be attributed to the 
  significant difference in the star formation rate at these 
 locations in NGC 300
 as shown in Sec.~\ref{Discussion} and Fig.~\ref{SFHsoln}.

\section{Star Formation History }\label{SFH} 

 In each of the 
 two fields at intermediate radii (F1 and F3)
 one finds that only the brightest stars have been 
 detected. Accordingly, the associated CMDs
 do not permit a  detailed derivation of the star formation
  history (SFH) using  synthetic CMDs.
 Despite this, we may  broadly  estimate how the heavy element abundances 
 and star formation rate (SFR) has changed during the lifetime of NGC 300.  
We have followed the method and have adopted the hypotheses introduced
 by Aparicio et al. (1997) and Mart\'{\i}nez-Delgado, 
 Aparicio \& Gallart (1999). We refer to these papers for a 
 detailed description of the method. However, 
 a brief summary here is appropriate. 

In short, the SFH is considered to be composed of three functions: the SFR,
$\psi$(t); the Initial Mass Function (IMF); and the chemical enrichment law, 
 Z(t). With these functions as input, the synthetic CMDs were 
 computed using scripts
  introduced by Aparicio et al. (1997). 
For the analysis, we have assumed a Salpeter IMF (Salpeter 1955) with low and
high mass cut-offs  at 0.7\,M$_\odot$ and 30\,M\,$_\odot$ respectively,
  and have ignored the possible effects of binary stars because the old
  population in our CMD is mainly composed by RGB  and AGB stars. Such stars
  are expected  to be well-mixed in the disk by the present time:  with a
  random velocity of 1\,km\,s$^{-1}$, the crossing time for a 1.5\,kpc region or
  the  WFPC2 field size would be about 1.5\,Gyr. 
 Consequently, Z(t) and the SFR have been determined for each 
 WFPC2 field rather than for the individual chips in order to maximize
  signal-to-noise.

For determination of Z(t),  we considered six age intervals
 with widths suited to the present use of  V-, and I-band photometry, 
 which is anyway relatively insensitive to  variations in Z(t).  These intervals are: 
 15 to 12\,Gyr, 12 to 9\,Gyr, 9 to 6\,Gyr, 6 to 3\,Gyr, 3 to
 1\,Gyr.   For  each age interval,  we adopted a constant SFR, and 
  synthesized a stellar
 population of 40,000 stars with absolute magnitudes brighter than
   M$_{\rm I}$ 
 = -3.0\,mag.   The chemical abundance (Z) of each star has been taken at random.  
 Internal photometric errors from  the artificial 
 star tests described in Sec.~\ref{obs_datared} have been included. 

 In order to estimate  the mean heavy element (Z) abundances
  of each age interval, 
 we adopted the following recipe. Firstly, we
 selected  stars from the synthesized population of stars  with
 heavy element abundances in an  arbitrary range (Z$_1$ to Z$_2$).
 Then  we measured the position and  full width of the 
  synthetic RGB/AGB locus for stars at 22.55 $<$ I (mag) 
 $<$ 26 and V-I $>$ 1.4\,mag.
 This procedure was repeated  a few hundred times for different Z ranges.
  From the  resultant  database, only  
 those  Z pairs  giving
  synthetic colours and widths that closely
 matched the values from the real CMD data were chosen.  In this way
 we could obtain an estimate of the maximum and minimum Z values in each age
 interval that are consistent with the real CMD data.
 The two 
 Z indicators (width and position of the RGB/AGB locus), however, 
  sometimes returned different Z pairs: the standard deviation
 of the differences  is taken as an estimate of the uncertainty in  Z(t) which
 is plotted in Fig.~\ref{Zlaw}.  For a test of the robustness of this
 Z(t) solution to the  selected
 CMD area, defined above,  we  varied the I-band range and colour
   boundary  by up to 0.1\,mag. The variation in Z is of
  the order of 20\% which is consistent with the largest error bar size.

\begin{figure}[htp]
 \includegraphics[width=15cm,height=12cm]{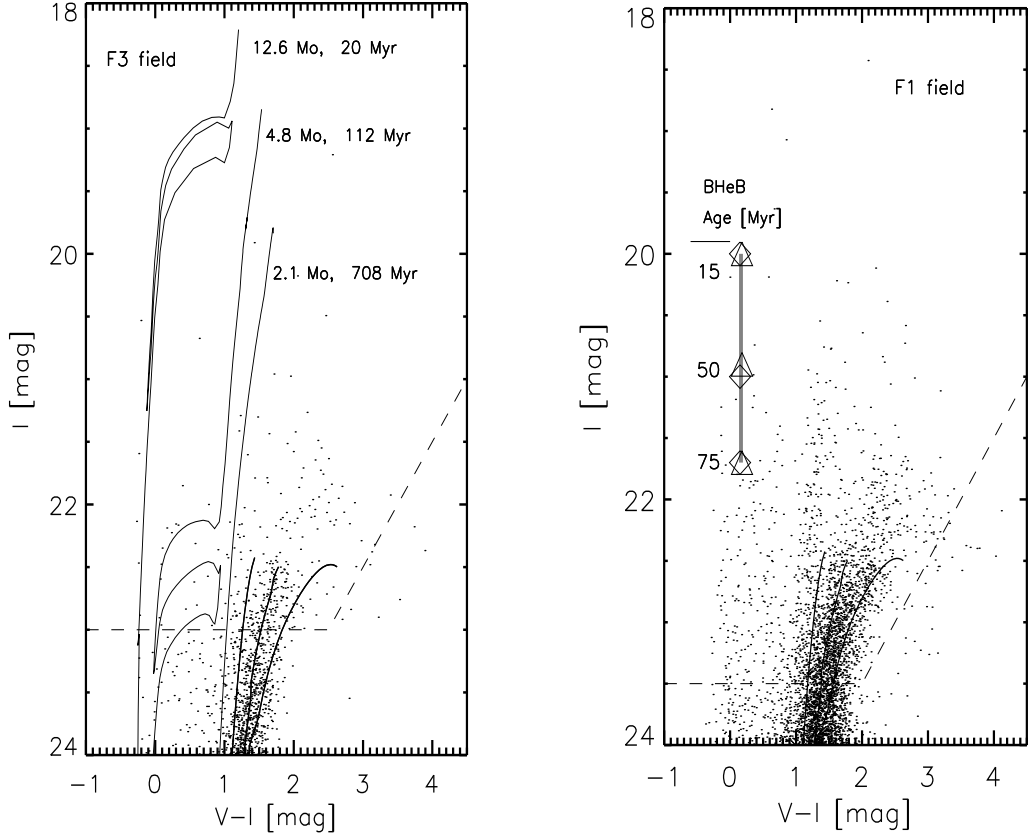}
 \caption{CMD for NGC 300 F1 and F3 fields.
 The stellar evolution tracks/isochrones
  for the adopted Z = 0.004 Padova models (see
  text) are  plotted ({\it thin/solid}) in the left panel.
  These are labelled with the
  starting mass in solar  units, and corresponding ages. 
 The  age -- I-band magnitude trend for blue supergiant/core 
 helium burning stars is marked inthe right-hand panel 
 for two metallicities, namely Z = 0.003 (diamonds) and Z= 0.005 (triangles);
  and age error  sizes are smaller than the  symbol sizes. 
  From metal-rich (right) to metal poor (left), reddened
 ridgelines ({\it thick/solid}) are for 47 Tuc, [Fe/H] = -0.71; NGC 1851, 
 [Fe/H] = -1.29;  NGC 6397, [Fe/H] = -1.91.
 The adopted true distance modulus is 26.5\,mag. The 90\% completeness
 limit is also indicated ({\it dashed}).
} \label{FigVIcmd_isoc}%
    \end{figure}

\begin{figure}[htp] 
\includegraphics[width=15cm,height=15cm,angle=-90]{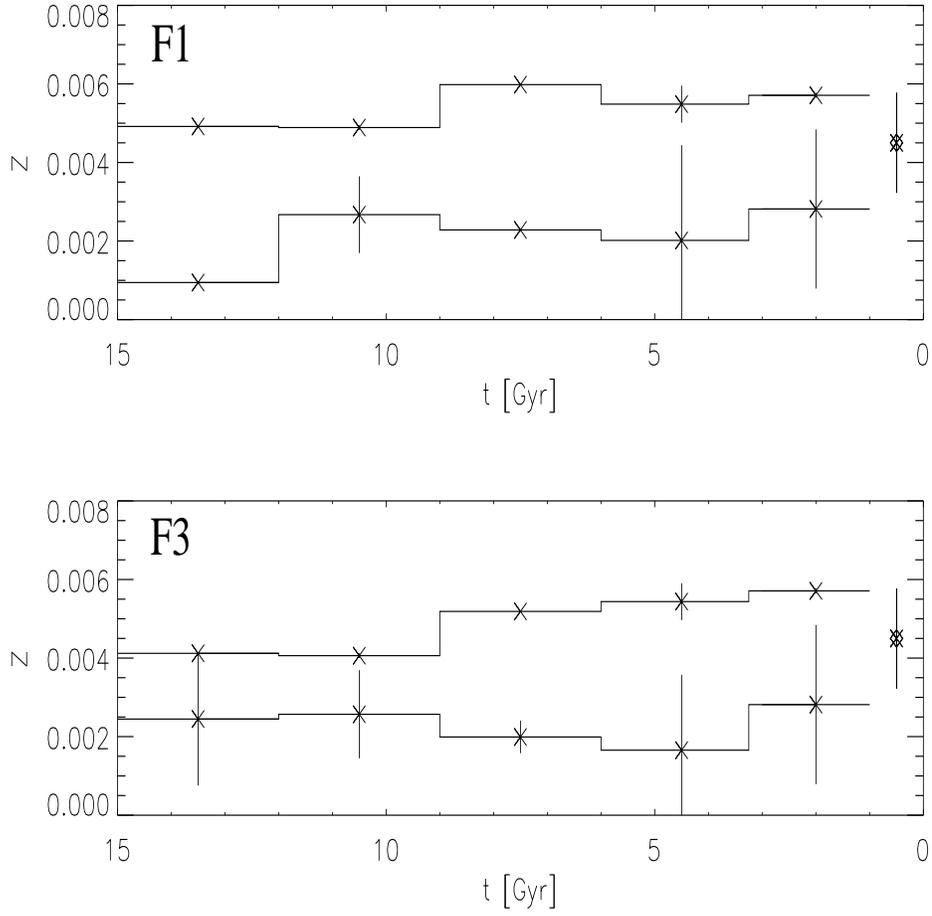}
   \caption{The derived variation in heavy element content 
 during the lifetime of  NGC 300 for two fields at different radial 
 distances from the galaxy centre. 
 See text in Sec.~\ref{SFH} for further details.}
   \label{Zlaw}
    \end{figure}

 For the past 1\,Gyr, 
  our only estimate  of  the  mean Z abundance  comes from
  spectroscopy of two  blue super-giants  
  outside the central region (r$\ga$ 2.3\arcmin)
  of NGC 300 (Bresolin et al. 2002b,  
  Urbaneja et al. 2002)\footnote{The CMD may not be used mainly
 due to the lack of a relatively Z sensitive feature for stellar ages below 
 about 1\,Gyr -- e.g. Fig.~\ref{FigVIcmd_isoc}, right-hand panel}. We take the average and rms of these, 
 and included them in Fig.~\ref{Zlaw}.
    It turns out that extrapolation of Z(t) to ages less than
 1\,Gyr leads to a fine agreement with the empirical estimates; but ongoing
   (e.g. Bresolin et al. 2002b)  and future
 spectroscopic studies  are needed for  better global Z information on NGC 300.

 Next, we computed a synthetic CMD for a stellar population 
 with ages in a narrow interval using a library of stellar evolution tracks 
 (Bertelli et al. 1994). We call this a partial CMD: taking a set of  
 partial CMDs, 
 each covering different age intervals (e.g. several times 10$^8$\,yr to a few
 times 10$^{9}$\,yr), the full galaxy age  ($\sim$ 15\,Gyr) is accounted
 for. A combination of partial CMDs that 
 covers the full age of the galaxy is called a candidate global synthetic CMD.

\begin{figure}[htp]
\includegraphics[width=15cm]{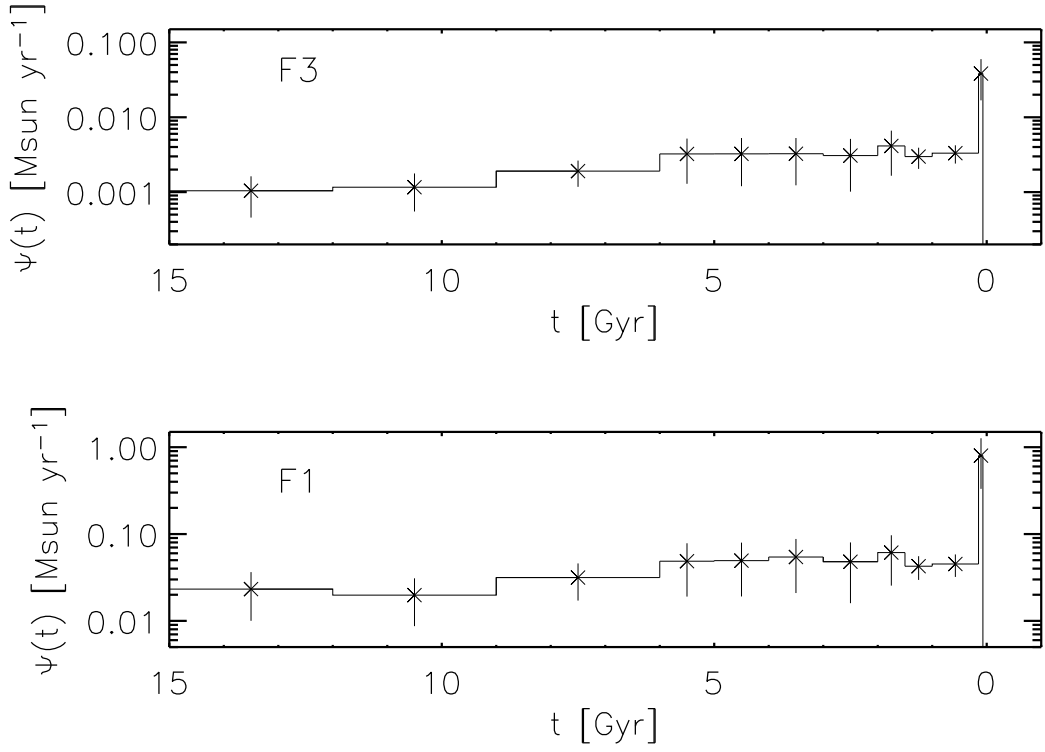}
   \caption{SFR solution for two NGC 300 fields using the partial model
     method, N6E1 (bottom) and centre-off (top). There is a significant
 star formation rate difference between the two fields.  
  See text in Sec.~\ref{SFH} for further details.}
   \label{SFHsoln}
    \end{figure}

For each real CMD, we computed a total of 50,000 candidate 
 global synthetic CMDs  in the following way.
  Each such CMD was made by  
 simply extracting all stars from the synthetized stellar population
  (mentioned earlier) such that in each age interval\footnote{Shorter age intervals were adopted for the past 6\,Gyr than for
 the Z(t) estimate  (see Fig.~\ref{SFHsoln}) because time resolution is
 potentially  much better for SFR calculations
 than that of Z over the past few $\times$10$^9$\,yr, especially if dramatic
 SFR changes could have occurred (e.g. see Williams 2002). 
 Interpolation was used to assign minimum and maximum Z values to the
 shorter age bins at ages greater than 1\,Gyr.}
  there are only stars of the right age and  
  each star has a synthetised Z value 
 between the estimated maximum and minimum
  value for that age interval. 
    In the next step, we selected eleven relatively age-sensitive  
  CMD boxes sampling the main features of the real CMD; 
 and counted the stars in those boxes (see Fig.~\ref{Star_XY_cmd_boxes}). 
 To paraphrase Aparicio et al. (1997), let
  us label the number of stars in box j of 
 the observed CMD with N$^{\rm o}_{\rm j}$, 
 and the number of stars in box j of the partial CMD i with 
  N$^{\rm m}_{\rm ji}$  (the partial CMD of the i$^{\rm th}$ age
 interval). m is the total number of candidate synthetic CMDs generated.
 The star counts per box are related to each other  by 

\begin{equation}
\rm N_{\rm j}^{\rm m} = k \sum_{\rm i} \alpha_{\rm i} N_{\rm ji}^{\rm m}
\end{equation}

 where   $\alpha_{\rm i}$ is the linear combination coefficient in age
 interval i,  and   k is a constant scaling factor.
  $\alpha_{\rm i}$  and k are related to the SFR, $\rm \psi(t)$, by:    

\begin{equation}
\rm \psi(t) = k \sum_{\rm i} \alpha_{\rm i} \psi_{\rm i} \Delta_{\rm i}(t)
\end{equation}

 where   $\Delta_{\rm i}$(t) = 1 if t is 
 inside the current age interval of
interest (corresponding to  the partial model i; or similarly, the model associated with the
 age interval i), and  $\Delta_{\rm i}$(t) = 0
 otherwise  (Aparicio et al. 2001); and $\psi_{\rm i}$  corresponds to partial
 model i.
  The $\rm \psi$(t) values having the highest probability of fitting
 the data in a chi-squared\footnote{$\chi^2$ = $\sum_{\rm k}$ 
 ((N$_{\rm j}^{\rm o}$ - N$_{\rm j}^{\rm fit}$) / N$_{\rm j}^{\rm o}$
where k is the number of candidate solutions.} sense can be 
 obtained by a least-squares fitting of  N$_{\rm j}^{\rm m}$ (i.e. simulated
 star counts)  to N$_{\rm j}^{\rm o}$ (i.e. observed or real star counts), where the $\alpha_{\rm i}$
 coefficients are the free parameters and  each is
  chosen  randomly in the range 0
 to 1.  
 Coefficients that provide synthetic CMDs that best match the real CMD
  will have the smallest   $\chi^2$ values. Accordingly,
 after producing 50,000 sets of coefficients (i.e. 50,000
 candidate SFR solutions\footnote{The common SFR zero point is determined by knowing the total initial 
 stellar mass of all stars generated while producing the 
  synthesized stellar population.}), one has a distribution 
 of $\chi^2$ values. The wide spread actually occurs because 
 there are no unique age indicators
 for each age interval in our CMDs: indeed, none are expected
  because of the 
 large degeneracy between age and chemical composition in our data\footnote{Such
 degeneracy would be partially broken by a longer colour baseline 
 than V-I, e.g. by  including suitable 
 near-infrared data. Importantly, however, star formation histories can be 
 measured accurately when photometry reaches to
 M$_{\rm V}$=+2\,mag (Dolphin 2002), or fainter still, to
  ancient main sequence turn-off stars.}. 
 Thus, we take the standard approach
 of taking the average of the best solutions with their
 standard deviation representing the uncertainty: here, we 
 define the best solutions as those meeting a 
   $\chi^2$ cut-off criterion. 
  This criterion is taken as  $\chi^2$ $<$  ($\bar{\chi}$$^2$ - 3$\sigma$), where $\sigma$
 is the standard deviation in the $\chi^2$ distribution.  With this,
 a few hundred SFR solutions or less than 
 1\% of all the candidate solutions were selected.
   For each  set of coefficients that satisfied the criterion,
  an estimate of the SFR(t)  was obtained; and the standard deviation 
 of all the SFR estimates in a given time interval
 was taken as the uncertainty in that interval.
 The mean SFR is plotted on a logarithmic scale
 against time in Fig.~\ref{SFHsoln} 
 for each field together with the estimated SFR uncertainty.

 In order to test the robustness of the derived SFR trend, we also 
 used a different cut-off, namely the   $\bar{\chi}$$^2$ - 2$\sigma$, 
 giving about 1700 global solutions, and  
 has the effect of increasing SFR error bars only slightly
 and has a negligible change on SFR(t). We note that the typical 
 error bar size indicates an uncertainty of  a factor of 
 about two to three in SFR in general, consistent with the 
 uncertainties presented in studies
 that considered the upper RGB data and 
  several age intervals between 15\,Gyr and now (e.g. Aparicio
 et al. (1997), their Fig. 8 -- LGS 3;  Dolplin (2002), his Fig. 7, panel e --
 simulated galaxy with a relatively complex  SFH). 
  It is evident that error size in Fig.~\ref{SFHsoln} tends to grow 
 towards younger ages: on one hand this 
 is due to the decreasing number of recently formed 
 stars  in the observed fields --
 which forces the use of wide age intervals for signal-to-noise reasons,
  while on the other hand  there might well have been episodic star formation
 during the past few $\times$10$^9$\,yr  
to which our age binning is possibly  fairly insensitive\footnote{Dolphin
  (2002; his Figs. 5 and 7) demonstrates this latter point through estimations 
 of the star formation histories of synthetic galaxies with episodic star
 formation histories.}.

 Another point concerns the ability to estimate both Z(t) and SFH(t)
 from  RGB and AGB star data alone, as opposed to also having data on 
 horizontal branch and fainter structure. The concern stems from the 
 difficulty of disentangling the contribution of age and chemical composition 
 to stellar colour, the so-called degeneracy between the two parameters.
  Thus, in such a situation solving simultaneously for the true 
 Z(t) and SFR(t) is difficult, even if possible, as the
 reasoning is flawed. The approach that we have applied is, however,
 one solution to this problem. 
  We recall that the approach that we have adopted has been to first
 assume a constant SFH and then to solve for Z(t). Then this Z(t) has then 
 been  used as input in the effort to obtain 
 the solution for the SFR(t). Given the
 uncertainties
 in the final Z(t) and SFH(t), the approach we have followed is deemed to be 
 tolerable.

 Even though a detailed chronology
 of  star formation in the WFPC2 field over the past few $\times$10$^9$\,yr
 is beyond the scope of the present paper, we can form 
 broadly acceptable conclusions on SF activity.
  Discussion of this issue is given in   Sec.~\ref{Discussion}.

\begin{figure}[htp]
\includegraphics[width=12cm,height=15cm,angle=-90]{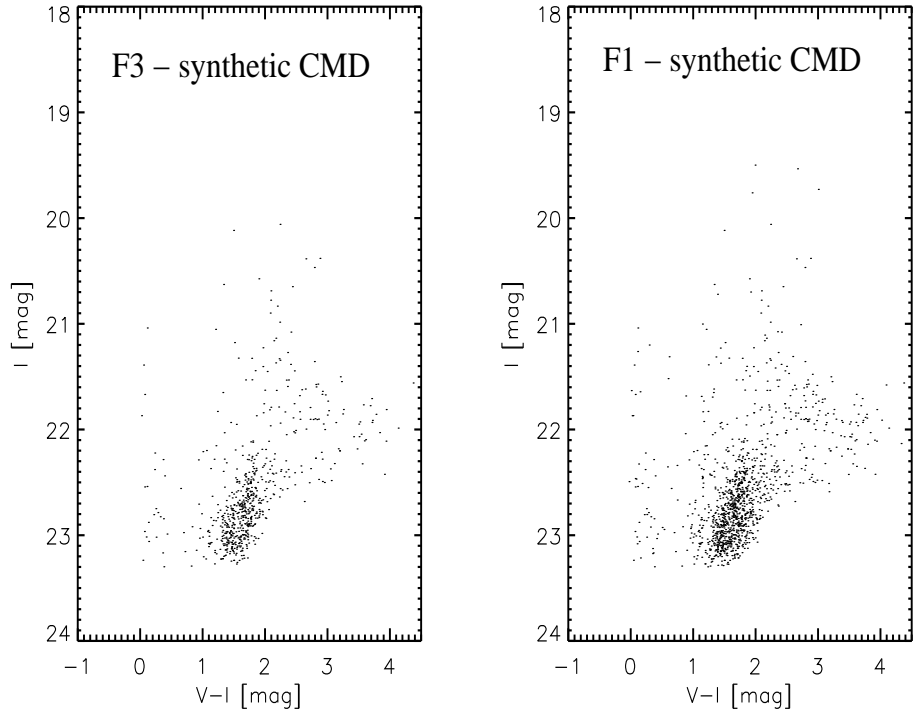}
  \caption{Synthetic CMDs for  two NGC 300 fields based on the SFH solution
 in Fig.~\ref{SFHsoln}. The CMD has been extended to I$>$23\,mag, for the
 sake of presentation, even though
 the SFR has been determined using stars brighter than this.
  See text in Sec.~\ref{SFH} for further details. }
   \label{synCMDs}
    \end{figure}

In order to test the reliability of the derived star formation history, we
 present the synthetic CMDs derived for the fields (F1 and F3) for 
 which an SFH
 analysis has been possible  (Fig.~\ref{synCMDs}), for comparison with the 
 real CMDs   (see Fig.~\ref{FigVIcmd_isoc}); photometry uncertainties in the
 real data have been simulated and added.  
     Overall, there is good agreement 
 for each  synthetic/observed CMD pair.  
   The main difference is in the colour--magnitude spread: 
 synthetic CMDs are inclined to have  tighter
 features than their counterparts in the real astronomical data.
 This could be due to a range of extinction values for the stars 
 (Williams 2002) that has not been estimated by the stellar 
 population synthesis models.
  We note the
  difference in the star count density at red colours V-I$\ga$  3\,mag 
  at/near the RGB peak magnitude in the F1 field, for which 
 the effect is pronounced, is not  unexpected
 because incompleteness becomes  significant for such stars.
  Additionally, 
 the required completeness correction increases significantly at 
 faint magnitudes (I $>$ 23\,mag) which causes significant
  scatter in the real data.

\section{Discussion -- Star Formation} \label{Discussion}

In Sec.~\ref{SFH}, we  determined a low time resolution 
 chemical enrichment law and SFR(t)
 for two disk fields in NGC 300. 
 In this Section, we summarize briefly the key findings based on WFPC2 data that give insight on the SFH of NGC 300.

The metallicity has been measured in two disk fields to be more metal poor on
 average  than 0.006 (or 0.33\,Z$_{\odot}$) during the lifetime of NGC 300
 and,  based on the available evidence, Z(t) appears to have
  changed relatively little during that time. 
 The 
 present day  value of Z may be of the order of 0.0035 to 0.0055 
  based on two empirical (spectroscopic) estimates from 
 Urbaneja et  al. (2003). Their best-fit oxygen abundance 
 (12 + log\, O/H) estimates for two blue supergiants are
  8.3 and 8.65\,dex  (their Table 1). 
 We note that Pagel et al. (1979)  measured 
 oxygen abundances  in six H$_{\rm II}$ 
 regions  in NGC 300. They found oxygen abundances approximately
 in the range 8.6 to 9\,dex.
 It is encouraging that the oxygen abundances
 from the two spectroscopy studies roughly agree; 
 this fact strengthens confidence in the derived values 
 of Z in the present paper for the past few $\times$10$^9$\,yr.

 For the star formation rate in  disk fields F1 and F3, 
 a simple estimate  has been made  (Fig.~\ref{SFHsoln}): 
 the derived mean SF rates are 0.002 and 0.04\,$\msun$\,yr$^{-1}$ in the F3
 and F1 fields respectively.  We note that the apparent SFR increase at
 about 100-200\,Myr may not be real because of the estimated uncertainty --
 the occurence of this feature could be reflecting a common
  systematic error for both field, possibly due to the small number statistics 
 available for such young stars, that is not assessed by the $\chi$-squared
 tests.  The  mean SF densities,  $\bar{\Psi}$/A,  are
     3 $\times$10$^{-4}$$\msun$\,yr$^{-1}$\,kpc$^{-2}$ and
   6 $\times$10$^{-3}$$\msun$\,yr$^{-1}$\,kpc$^{-2}$  in
the F3 and F1 fields respectively.
 These SF density values are consistent 
 with expectations for spiral disks:
 the SFR density in spiral disks can be up to 
 about 0.1\,M$_\odot$\,yr$^{-1}$\,kpc$^{-2}$ in
 spiral disks and typical activity exhibits a steady state behaviour
 (Kennicutt 1998);  SF in nearby spirals has, on average, 
  a strong dependence on galaxy type, 
 weak or no dependence on spiral structure or the presence of a bar, and
 moderate dependence on past interactions  (Kennicutt 1998).
      It may well be that the average
  SF activity in the NGC 300 disk follows the typical
  trend in nearby spiral disks,  but conclusive evidence will 
 only be obtained from future observations covering more of the galaxy.

\begin{figure}[htp]
  \centering{\includegraphics[width=15cm]{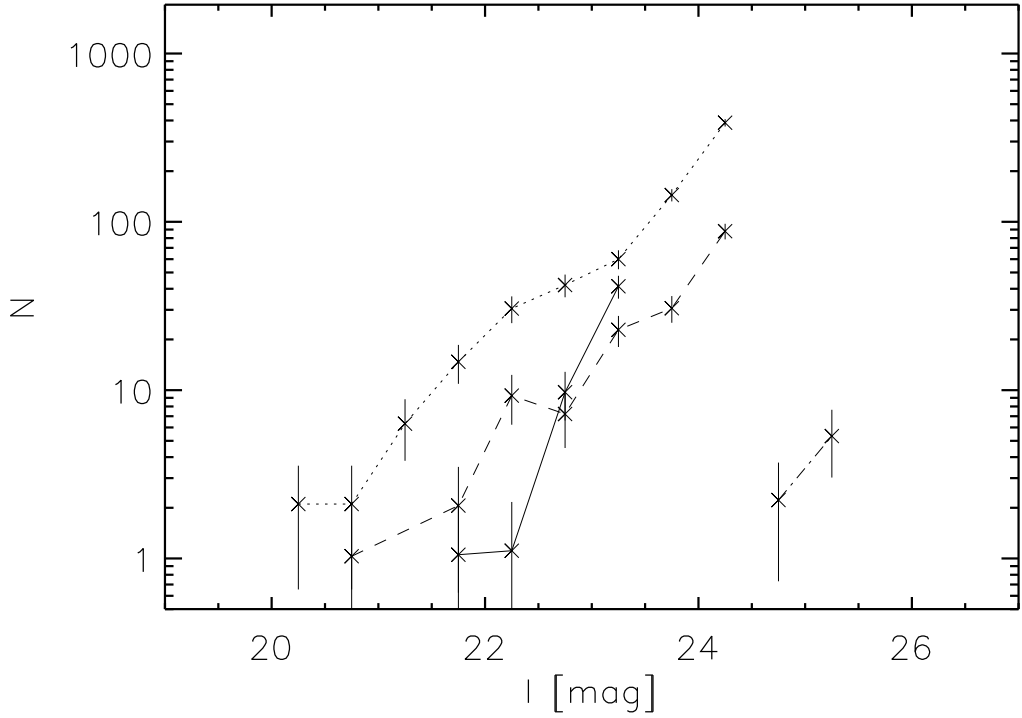}
}
   \caption{I-band luminosity functions for four WFPC2 fields, corrected for
     incompleteness, for stars above the 50\% completeness threshold with
 V-I $<$ 0.6 (mag).  
 These  fields are F1 ({\it dotted}),  F2 ({\it solid}), F3 ({\it dashed}), and F4
 ({\it dot-dashed}).  Error bars are $\pm$$\sqrt{\rm N}$. 
  See text in sub-Sec.~\ref{Discussion} for further details.}
   \label{Ilumfn}
    \end{figure}

 On the other hand, Davidge (1998) reported a suppression of recent star formation in the central 
 region of NGC 300 based on near-IR observations. 
   In our completeness-corrected I-band luminosity 
 functions  for stars  younger than a few times 10$^8$\,yr (at V-I $<$ 0.6\,mag),
 and  brighter than the  50\% completeness 
 limits,
 we also find a  significant cut-off in the central function
 with respect to the F1 and F3 fields (Fig.~\ref{Ilumfn}). 
 However, as mentioned in Sec.~\ref{dust}, the 
 central extinction is unknown and may be significant: from the ESO/WFI image
 there is evidence for distinct dust shells in part of the F1 FoV that are
 superimposed
 on  a much more widely spread dust cloud. Consequently, 
 internal extinction  could 
 possibly explain the truncation of the central I-band luminosity function:
 about one to two magnitudes of I-band 
 extinction would be 
 needed in order to dim the  brightest object 
 observed in the outer fields (I$\sim$18\,mag) to  match the brightest stars
 in the central field (Fig.~\ref{FigVIcmds}). 
  Lastly, central ring-like gas (but also dust) structures are expected
 based on simuations of the  detailed chemical and dynamical 
 evolution  of galaxies  (e.g. Samland \& Gerhard 2003).

 Another issue is that there has been an 
 increase in SFR over the last few Gyr in NGC 300. 
 We know that 
 the observed increase in SF activity
 in the  F1 and F3 disk fields is not unprecedented in some spiral
 galaxy studies: some sites in 
 M31 have SF rates indicating that  sufficient gas may have been
  retained and/or
 acquired for continued or episodic star formation (Williams 2002; Williams
 2003, based on the SFR measured in OB associations).

 Finally, there is the issue of why there has been relatively little increase
 in Z over  time in NGC 300 and why it appears not to have changed 
 significantly over the lifetime of the galaxy. A possible 
 solution is that metal-poor gas has been acquired by NGC 300.
 In the context of metal-poor gas acquisition 
 through  past encounters, we note that 
 intergalactic gas and stars may be a partial signature of 
  such  encounters: a 
  clustering of  H$_{\rm I}$ clouds around NGC 300 and NGC 55 was found
  by Haynes \&  Roberts (1979) based on  sensitive 21\,cm observations; but uncertainty
 over group membership of clouds remains (Haynes \&  Roberts 1979).
 We also  note that colour gradients and morphological studies of 
 spiral galaxies can reveal clues about 
 past mergers: no evidence for a colour gradient in NGC 300 was found by
 C85, and  a detailed study of the morphology
 of NGC 300  (e.g. examining distorted isophotes hinted at in
  Fig.~\ref{DSS2red_NGC300_Chart}) remains to be tackled.

\section{Conclusions} \label{Conclusions}
We have presented the  first  WFPC2 V, I photometry for the 
Sculptor Group late-type spiral galaxy NGC
300 in four fields ranging from the  centre to the outer-edge.
  In particular, we have the following conclusions.

 \noindent 1. We have derived the first estimate of the 
 star formation history in two disk  fields. 
 Our analysis indicates that  stars were born at similar
 epochs and star formation activity possibly increased on average over time,
  but at  different mean rates.

  \noindent 2. The main stellar population is predominantly old, 
 consisting of  RGB and AGB stars, 
 based on a synthetic CMD analysis.
  Z is found to have been more metal poor than  0.006 (or 0.33\,Z$_{\odot}$) 
 with no evidence for significant change in the mean Z value
  over time in both disk  fields.
 
 \noindent 3.  In the circum-nuclear region, we find a dearth of bright stars 
   relative to two disk fields that cannot be explained
 by observational effects. Taken at face value, this finding would agree
 with the Davidge (1998) report of 
 suppressed star formation there during the past 10$^9$\,yr; but the
  possibility  of significant 
 central I-band extinction (up to about 1-2\,mag) remains. 

\noindent 4. We have also reported a newly detected a  young star association
 (probably up to $\sim$ 10\,Myr old)   of about average size ($\sim$140\,pc)
  in one of the disk fields that probably 
 coincides with a star-forming region. 

\noindent 5. Lastly, the distance modulus has been determined using 
  Ferrarese et al. (2000) Cepheid-based calibration the tip magnitude of
 the RGB to be (m-M)$_\circ$ = 26.56$\pm$0.07 
 ($\pm$0.13)\,mag. This is in good  agreement with the Cepheid distance
 to NGC 300 from Freedman et al. (2001).
 Incidentally, both values are in good agreement between with the 
 TRGB distance estimated using the Lee et al. (1993) empirical calibration, 
 which is Cepheid independent. 
 There is, however, a discrepancy 
 with the theoretical calibration of the RGB tip magnitude
 from SC98, which may be a zero-point
  problem. Accordingly, this 
 means that the same difference 
 in magnitude between the Cepheid distance and the theoretical 
 calibration of the distance for 
 NGC 300 exists for galaxies in the 
 HST key project on the extragalactic distance scale.

\acknowledgments
 A. Dolphin is thanked for help using HSTphot, as is A. Aparicio for
  useful comments on the population synthesis code.  It is a pleasure 
 to thank the anonymous referee for several useful questions and comments. 
  DB acknowledges the support of the European research and training network on
 Adaptive Optics for Extremely Large Telescopes under contract
 HPRN-CT-2000-00147.

\clearpage

\clearpage

\begin{deluxetable}{llllccc}
\tabletypesize{\scriptsize}
\tablecaption{Source list: NGC 300 HST/WFPC2 archival data}
 \tablewidth{0pt}
\tablehead{
  \colhead{Field} &   \colhead{Radial} & \colhead{Radial}   & \colhead{Date}  & \colhead{Filter}  & \colhead{Number of} &  \colhead{Integration Time} \cr
 &  \colhead{distance$^a$}     & \colhead{distance$^b$}         &    &
 \colhead{}  &   \colhead{Frames} \cr
  &  \colhead{(arc min)}     & \colhead{(kpc)}    &    &   &  & \colhead{(s)}
}
\startdata
 \rm F2 & 0.44  & 0.4 & \rm 2001\,May\,06 &  \rm F814W/F547M & \rm 2/4  & 300/400 \\
 \rm F1 & 5.99  & 5.1 & \rm  2001\,July\,02 &  \rm F814W/F606W  & \rm 2/2 & 300/300  \\
 \rm F3 & 7.12  & 6.0 &  \rm 2001\,Sept.\,13 &  \rm F814W/F555W & \rm 2/2  & 500/500 \\
 \rm F4 & 12.84 & 10.9 & \rm  2001\,June\,20 &  \rm F814W/F606W  & \rm 4/4 & 500/500 \\
 \enddata
\tablenotetext{a}{Separations from cluster centre to the coordinates listed 
 in the data pointings table (nominally the middle of the field of view)
  are taken from  MAST, the
  multimission archive at STScI.}
\tablenotetext{b}{De-projected radial distance in kpc based on 
 an inclination  of 46$^\circ$ (Tully 1988).}\label{obs_log}
 \end{deluxetable}

\clearpage

\begin{deluxetable}{lll}
\tabletypesize{\scriptsize}
\tablecaption{Star counts in selected regions of the F1 and F3 CMDs.}
 \tablewidth{0pt}
\tablehead{   \colhead{Star counts} & \colhead{F1}   & \colhead{F3} \cr
  &  \colhead{(arc min)}     & \colhead{(arc min)}   }
%} 
 \startdata
\rm D: {\rm old\, AGB,\, RGB}  &   \rm  646 (4.1\%)  &  \rm  200 (7.5\%)   \\
\rm C: {\rm old\, AGB}   &   \rm 173 (8\%) &  \rm   49 (15\%)   \\
\rm B: {\rm young\, AGB}  &  \rm  74 (12.2\%) &  \rm  23 (22\%)  \\
\rm A: {\rm $<$ \, 1\,Gyr}  & \rm  330 (5.8\%) &  \rm   68  (13\%)  \\
\hline
\rm Count \,\, ratio  & &   \\
\rm C/D & \rm   0.27 (0.04)  &  \rm  0.24 (0.07) \\
\rm B/D & \rm   0.12 (0.01)    &  \rm  0.12 (0.03) \\
\rm A/D  &  \rm   0.51 (0.05) & \rm  0.34 (0.07) \\
  \enddata
\\
\noindent {The  star counts (N) and ratios (corrected for a mean incompleteness of 10\% at 
 I $<$ 23\,mag)   come from  defined CMD boxes (see
   Fig.~\ref{Star_XY_cmd_boxes}).    Star count error
  is $\pm$ $\sqrt{\rm N}$,  
 determined for each uncorrected star count.  
  Percentage error (top panel) and actual error (bottom panel)
 are given in parentheses. 
  Box boundaries are defined in Fig.~\ref{Star_XY_cmd_boxes}. }\label{table_F1F3counts}
 \end{deluxetable}
\clearpage

\clearpage

\clearpage

\clearpage

\clearpage

\begin{deluxetable}{lll}
\tabletypesize{\scriptsize}
\tablecaption{Coordinates$^{a}$ and size of newly detected stellar  association.}
 \tablewidth{0pt} \tablehead{  \colhead{$\alpha$} & \colhead{$\delta$}   & \colhead{FWHM} \cr 
   \colhead{(J2000.0)} & \colhead{(J2000.0)} & \colhead{(pc)}  
}
\startdata
  \rm  00\,54\,56.6 & -\,37\,35\,25.0 & 130 - 145 \\
\enddata
\tablenotetext{a}{Units of right ascension are hours, minutes, and seconds,
and units of declination are degrees, arcminutes, and arcseconds. 
Coordinates are taken from the online Aladin sky atlas: position error is
estimated to be 10-20$^{\prime\prime}$. }\label{table_newAssociation}
 \end{deluxetable}

\clearpage

\clearpage

\clearpage
\begin{deluxetable}{llllll}
\tabletypesize{\scriptsize}
  \tablecaption{Distance-related information for NGC 300.}
 \tablewidth{0pt}
  \tablehead{\colhead{Field} & \colhead{I$_{\rm TRGB}$}   & 
 \colhead{M$_{\rm I, TRGB}$ $\pm$ r $\pm$ s}  & \colhead{A$_{\rm  V}$ } & \colhead{A$_{\rm I}$} & \colhead{(m-M)$_\circ$ $\pm$ \rm r $\pm$ \rm s}  \\
  \colhead{($^{\prime}$)} &  \colhead{(mag)} & \colhead{(mag)} &
  \colhead{(mag)}   & \colhead {(mag)} \\
  \colhead{(1)} & \colhead{(2)}  & \colhead{(3)} & 
\colhead{(4)}  & \colhead{(5)}  & \colhead{(6)} }
\startdata
 \rm F1 & \rm 22.52 $\pm$ 0.02  & -4.06 $\pm$ 0.07 $\pm$ 0.13$^a$ & 0.039  &    0.019   & 26.56  $\pm$ 0.07 $\pm$ 0.13 \\
 \enddata
\tablenotetext{a}{Ferrarese et al. (2000), with r = random and s = systematic.}
\tablenotetext{b}{Taking median value, as a Gaussian fit is not possible; 
Columns (1) Radial distance of WFPC2 field from galaxy centre; (2) Tip I-band
magnitude of the RGB. (3) Absolute I-band magnitude of the RGB tip; 
(4) and (5) Galactic foreground extinction (Schlegel,
Finkbeiner,  \& Davies 1998) converted to I-band  using the Cardelli, Clayton,
 \& Mathis (1989) extinction law, and R$_{\rm V}$ = 3.1 (NED); (6) distance modulus}\label{dist_details}
\end{deluxetable}

\clearpage

\clearpage

\clearpage

\clearpage

\clearpage

%% You can append references to a table using the \tablerefs command.

%% Tables may also be prepared as separate files. See the accompanying
%% sample file table.tex for an example of an external table file.
%% To include an external file in your main document, use the \input
%% command. Uncomment the line below to include table.tex in this
%% sample file. (Note that you will need to comment out the \documentclass,
%% \begin{document}, and \end{document} commands from table.tex if you want
%% to include it in this document.)

%% \input{table}

%% The following command ends your manuscript. LaTeX will ignore any text
%% that appears after it.

\end{document}